\documentclass[preprint,
superscriptaddress,
amsmath,amssymb,aps,showkeys,showpacs,
twoside,final,secnumarabic,
nofootinbib]{revtex4-2}

\usepackage[paperwidth=205mm,paperheight=290mm,top=17mm,bottom=25mm,
inner=17mm,outer=17mm,
twoside]{geometry}

\usepackage{cmap} 
\usepackage[T1,T2A]{fontenc}
\usepackage[utf8]{inputenc}
\usepackage[russian,english]{babel}
\usepackage{color}
\usepackage{graphicx}
\usepackage{dcolumn}
\usepackage{bm} 
\usepackage[unicode=true,colorlinks=true,linkcolor=magenta, urlcolor=blue, citecolor = blue,breaklinks]{hyperref}
\usepackage{multirow}
\usepackage{url}
\usepackage{breakurl}
\DeclareGraphicsExtensions{.eps}

\newcount\issue
\newcount\Vol
\newcount\numb
\usepackage{fancyhdr} 
\def\Vol{\textbf{80}}
\def\numb{x}
\begin{document}

\title{
On  Lagrangian formulations for (ir)reducible mixed-antisymmetric \\ higher integer spin fields in Minkowski spaces}

\def\addressa{Center of Theoretical Physics,
Tomsk State Pedagogical University, 634061 Tomsk, Russia}
\def\addressb{National Research Tomsk State  University,
634050 Tomsk, Russia}
\def\addressc{National Research Tomsk Polytechnic   University,\\
634050 Tomsk, Russia}
\def\addressd{  IILM University, Knowledge Park II Greater Noida,\\
201306  Uttar Pradesh, India}

\author{\firstname{Alexander A.}~\surname{Reshetnyak}}
\email[E-mail: ]{reshetnyak@tpu.ru }
\affiliation{\addressa{},
\addressb{}, \addressc{}}
\author{\firstname{Julia V.}~\surname{Bogdanova}}
\affiliation{\addressc}
\author{\firstname{Vipul K.}~\surname{Pandey}}
\affiliation{\addressd}


\begin{abstract}
 We extend the results of Lagrangian formulations study  to construct  gauge-invariant Lagrangians  for (ir)reducible  integer
higher-spin massless and massive representations of the Poincare group with a
Young tableau $Y[\hat{s}_1,\hat{s}_2,\hat{s}_3]$ in  $d$-dimensional flat space-time (as the probable candidates to describe the Dark Matter problem beyond the SM). These particles are described   within a metric-like formulation by  tensor fields with 3 groups of antisymmetric Lorentz indices $\Phi_{\mu^1[{\hat{s}_1}],\mu^2[{\hat{s}_2}], \mu^3[{\hat{s}_3}]}$  on a basis of the
 BRST  method with complete, $Q$, and incomplete, $Q_c$, BRST operators.  We found unconstrained (with  $Q$) and constrained (with $Q_c$ and off-shell BRST invariant holonomic constraints) gauge Lagrangian formulations with different configuration spaces  and reducibility stages. The deformation procedure to construct interacting  gauge model with mixed-antisymmetric fields is proposed.
\end{abstract}

\par
\keywords{gauge invariance, BRST approach, cubic vertex, higher spin fields  \\[5pt]}

\maketitle
\thispagestyle{fancy}


\section{Introduction}\label{intro}

Problems of higher-spin (HS) field theory have been the subject of significant
discussions among theoretician community. Interest in HS field theory (for review, see e.g. \cite{reviews}, \cite{reviewsV}, \cite{Snowmass}, \cite{Ponomarev}) is
mainly inspired by hope to develop the new  Lagrangian
models in classical field theory and may be formulate on this
ground the new tools for the unification of the fundamental
interactions. HS field theory is closely related to
(super)string theory \cite{SFT}, \cite{tensionlessl} which operates with an infinite tower of bosonic
and fermionic HS fields in $10$-dimensional space-time with compactified extra dimensions thus generating theirs masses within dimensional reduction mechanism. Because of interacting (super)strings seems to be only unique way to consistently describe evolution of early Universe and
 involve the interaction of particles with different values of spin, starting from known particles with lower spins (Higgs particle for $s=0$; leptons and quarks for $s=\frac{1}{2}$; photon, $W^{\pm}, Z$ bosons for $\lambda=1$ and yet undiscovered graviton for $\lambda=2$) the unknown particles both with HS $s>2$  and with  generalized spin $s_{(k)}=(s_1,s_2,...,s_k)$ are also  admissible. From this perspective such particles  may be considered as alternative candidate to describe Dark Matter, see for a review\cite{DM1}, \cite{DM2} (always surrounding electromagnetically observable particles) beyond the extensions of the Standard Model involving sterile neutrino  \cite{Dubinin} and dark massive photon \cite{DMphoton}. Whereas the problem of Lagrangian formulation (LF) construction for HS fields in constant curvature spaces have been considered in different approaches (see for review, \cite{reviews3}, \cite{Bengtsson}) both in metric and frame-like formalism for totally, e.g. in  \cite{SinghHagen}, \cite{Fronsdint}, \cite{FronsdintA}, \cite{Franciatr2}, \cite{0702161}, \cite{0703049}
  and mixed-symmetric irreducible (spin)-tensors fields on $\mathbb{R}^{1,d-1}$ (or AdS space), e.g. in \cite{Labastida}, \cite{0311164}, \cite{0501108}, \cite{08070903}, \cite{08081778}, \cite{franciamixbos},
 \cite{1110.5044}, \cite{1211.1273}, \cite{2305.00142}  (for $d=4$ as well as for irreducible representations (irreps) with continuous spin, see  \cite{FedorukBuch}, \cite{BuchKrychcont}, \cite{2403.14446}, \cite{2503.14290}), the consistent interacting LF with HS fields is still far from its complete realization, in spite for the results (see, e.g., the papers and references therein) for cubic  \cite{frame-like1}, \cite{Metsaev0512}, \cite{BRST-BV3},  \cite{frame-like2},
 \cite{BRcub}, \cite{BRcubmass}, \cite{BRcubmass2}  and quartic \cite{DT}, \cite{Taronna} vertices.

 The another way to describe Poincare group $ISO(1,d-1)$ irreps, for $d>4$,  is to consider totally and mixed-antisymmetric (spin)-tensors fields, for   which the LF for the former  have been found in \cite{ashsb1}, \cite{ashsf2},     \cite{zinoviev} and for massless HS fields with antisymmetric spin ${s}{[2]}=(\hat{s}_1, \hat{s}_2)$ in \cite{0311212}, \cite{0402180}, \cite{Boulanger2}, \cite{zinoviev1} and in BRST (Becchi-Rouet-Stora-Tyutin) approaches with complete and incomplete BRST operators for massive and massless cases in \cite{resh1}.

 The universal BRST approach has the origin both from special supersymmetry, known as Becchi-Rouett-Stora-Tyutin symmetry \cite{BRST1}, \cite{BRST2} in Lagrangian quantization for the field theories with gauge group and from generalized canonical quantization, known as BFV (Batalin-Fradkin-Vilkovisky) \cite{BFV, BFV1, BFV2} method for quantizing constrained dynamical systems, but  serving to solve inverse problem of finding LF starting from non-Lagrangian partial differential equations which describe field (ir)reps of $ISO(1,d-1)$ or $SO(2,d-1)$ group. BRST approach, being also the realization of AKSZ model \cite{AKSZ},
  works with the superalgebra of operators $\{o_I\}$ containing together with isometry subalgebra, $\{o_A\}$, with only space-time derivatives the subalgebra with holonomic 2-nd class operator constraints $\{o_a, o^+_a\}$ \cite{}. In case of construction the nilpotent Grassmann-odd  BRST operator with respect to only subalgebra of $\{o_A\}$ it is named as incomplete,$Q_c$ following from SFT \cite{tensionlessl}, \cite{BRST-first}  in presence of holonomic constraints
\cite{0406192}, \cite{resh2}, as compared to the  complete one, $Q$ initiated by  A.~Pashnev, M.~Tsulaia, I.~Buchbinder  in  \cite{Pashnev1}, \cite{BurdikPashnev}, \cite{symint-adsmassless}
  \begin{equation}\label{QQc}
 Q_c = C^Ao_A+ \frac{1}{2}
    {C}^A{C}^Bf^D_{BA}{\mathcal{P}}_D (-1)^{\epsilon({p}_D) + \epsilon({o}_A)}, \ \ Q = C^Io_I+ \frac{1}{2}
    {C}^I{C}^Jf^K_{JI}{\mathcal{P}}_K (-1)^{\epsilon({o}_K) + \epsilon({o}_I)}
\end{equation}
with structural constants $f^K_{JI}(f^D_{BA})$ from involution relations: $[o^I, o_J\} = f^K_{IJ}o_K$, with operators of ghost coordinates $C^I$, momenta ${\mathcal{P}}_J$ of opposite to $o_I$ Grassmann parities and with $[C^I, {\mathcal{P}}_J\} =\delta^I_J$.

The first purpose of the paper is to  construct  general
gauge-invariant Lagrangians for both massless and massive
mixed-symmetry tensor fields of rank $\hat{s}_1 +\hat{s}_2+ \hat{s}_3$, with
any integer numbers $\hat{s}_1 \geq \hat{s}_2\geq \hat{s}_3 \geq 1$ for $\hat{s}_1
< [d/2]$ in $\mathbb{R}^{1.d-1}$, as
elements of Poincare-group $ISO(1,d-1)$ irreps with a Young
tableau having $3$ columns. Another aim is to formulate the deformation procedure to construct interacting  gauge model with mixed-antisymmetric fields. \\
\indent
In the Section~\ref{sec1:level1} we firstly construct LF with complete BRST operator for HS field. A LF with incomplete BRST operators and appropriate set of holonomic constraints is derived in Section~\ref{sec2:level1}. Section~\ref{sec3:level1} contains the concept of deformation procedure for  interacting LF of $p$ samples of irreducible HS fields subject to $Y[\hat{s}_1,\hat{s}_2,\hat{s}_3]$.\\
\indent
We use the definitions and notations  from
\cite{BRcub}, \cite{BRcubmass},  \cite{resh1}  for a metric tensor $ \eta_{\mu\nu}$ and  $(\epsilon, gh_H)(F)$,  $[\ ,\ \}$, $[x]$,
${s[3]}$, $({s}{[3]})_p$, $A^{[i}B^{j]k}$ for the values of Grassmann parity and   ghost number of a
 quantity $F$, supercommutator, the
integer part of
 real $x$, for  the  antisymmetric spin with components   $
(\hat{s}_1, \hat{s}_2,\hat{s}_3)$, for  the  $p$-multiple  $
({s}^1_{[3]},{s}^2_{[3]},...,{s}^p_{[3]})$ and for antisymmetrization
 $A^{[i}B^{j]k} =
A^{i}B^{jk}-
A^{j}B^{ik}$.

\section{\label{sec1:level1}Lagrangian formulation with complete BRST operator}

Here, we shortly recall \cite{resh1} that a massless (massive) integer spin irreducible
representation of Poincare group in a  Minkowski space
$\mathbb{R}^{1,d-1}$ which is described by a real-valued tensor field
$\Phi_{\mu^1[{\hat{s}_1}],\mu^2[{\hat{s}_2}], \mu^3[{\hat{s}_3}]}
\hspace{-0.2em}$ $\equiv $ $\hspace{-0.2em}
\Phi_{\mu^1_1\ldots\mu^1_{\hat{s}_1},...,\mu^3_1\ldots\mu^3_{\hat{s}_3}}$
 of rank $\sum_i^3\hat{s}_i$ and generalized spin
 $\mathbf{s} = (3,3,...,3,2,2,...,2,1$ $,...,1)$, ($\hat{s}_1\geq \hat{s}_2\geq \hat{s}_3>0, \hat{s}_1
\leq [d/2])$ to be corresponding to a Young tableau with $3$ columns
of height  $\hat{s}_1, \hat{s}_2,\hat{s}_3$, respectively (with omitting further the symbol "$\hat{\phantom{s}}$" under $\hat{s}_i$)
\begin{equation}\label{Young 3}
\Phi_{\mu^1[{s_1}],...,\mu^3[{s_3}]}
\hspace{-0.3em}\longleftrightarrow \hspace{-0.3em}
\begin{array}{|c|c|c|}\hline
  \!\mu^1_1 \!&\! \mu^2_1\! &\! \mu^3_1\! \\
    \cline{1-3} \!\cdot\! &   \!\cdot\!&   \!\cdot\!   \\
   \cline{1-3}
     \! \mu^1_{s_3}\! &\!\mu^2_{s_3}\! &\! \mu^3_{s_3}\!   \\
     \cline{1-3}
     \! \mu^1_{s_3+1}\! &\!\mu^2_{s_3+1}\! \!\\
       \cline{1-2} \!\cdot\! &   \!\cdot\! \\
   \cline{1-2}
     \! \mu^1_{s_{2}}\! &\!\mu^2_{s_{2}}\! \\
           \cline{1-2} \!\mu^1_{s_2+1}\!   \\
  \cline{1-1} \!\cdots\!\\
  \cline{1-1} \!\mu^1_{s_1}\! \\
  \cline{1-1}
\end{array}\ ,
\end{equation}
This field is antisymmetric with respect to the permutations of each
type of Lorentz indices
 $\mu^i$,
  and
obeys to the wave (Klein-Gordon) (\ref{Eq-0b}), divergentless
(\ref{Eq-1b}), traceless (\ref{Eq-2b}) and mixed-antisymmetry
equations (\ref{Eq-3b}) :
\begin{eqnarray}
\label{Eq-0b} &&
(\partial^\mu\partial_\mu (+ m^2))\Phi_{\mu^1{[s_1]},\mu^2[{s_2}],\mu^3[{s_3}]}
 =0, \\
&&    (\partial_{i} \Phi)_{
{\mu}^1[{s_1-\delta_{i1}}],...,{\mu}^3[{s_3-\delta_{i3}}]} \equiv  \partial^{\mu^i_{l_i}}\Phi_{\mu^1{[s_1]},\mu^2[{s_2}],\mu^3[{s_3}]} =0,\texttt{ for } 1 \leq l_i \leq s_i,\ i=1,2,3, \label{Eq-1b}
\\
&& (\texttt{Tr}^{ij}\Phi)_{\mu^1{[s_1]},\mu^2[{s_2}],\mu^3[{s_3}]}\equiv  \eta^{\mu^i_{l_i}\mu^j_{l_j}}\Phi_{\mu^1{[s_1]},\mu^2[{s_2}],\mu^3[{s_3}]}=0,  \ \ 1 \leq i<j\leq 3,\label{Eq-2b}\\
&& (Y^{ij} \Phi)_{
\mu^1[{s_1}],[\mu^i[{s_i}],...,\mu^j_{l_j}],\hat{\mu}^j[{s_j-1}],\mu^3[{s_3}]} \equiv  -(-1)^{l_j} \Phi_{
\mu^1[{s_1}],[\mu^i[{s_i}],\mu^j_{l_j}]\underbrace{\mu^j_1...\mu^j_{l_j-1}\mu^j_{l_j+1}}...\mu^j_{s_j},\mu^3[{s_3}]}=0, \label{Eq-3b}
\end{eqnarray}
with use the notations for the operations  of divergences $(\partial_{i} \Phi)$, mixed traces $(\texttt{Tr}^{ij}\Phi)$, Young-antisymmetrizations
$(Y^{ij} \Phi)$.
The  bracket below in (\ref{Eq-3b}) denote that the indices  in it do not
include in  antisymmetrization.

It is easy to see, the system (\ref{Eq-0b})--(\ref{Eq-3b}) is equivalent
to set of primary constraints on the string--like vector $|\Phi\rangle \in \mathcal{H}^f$ in an auxiliary Fock
space $\mathcal{H}^f$, generated by $3$ pairs of Grassmann-odd  (antisymmetric
basis) oscillators $\hat{a}^{i+}_{\mu^i}(x)$,
$\hat{a}^{j}_{\nu^j}(x)$, $\mu^i,\nu^j
=0,1...,d-1$:
\begin{eqnarray}
\label{PhysState}  \hspace{-2ex}&& \hspace{-2ex} |\Phi\rangle  =
\sum_{s_1=0}^{[d/2]}\sum_{s_2=0}^{s_1}\sum_{s_3=0}^{s_{2}}
\frac{\imath^{\sum_{p=1}^3s_p}}{{s_1!s_2!s_3!}}\Phi_{\mu^1{[s_1]},\mu^2[{s_2}],\mu^3[{s_3}]}\,
\prod_{i=1}^3\prod_{l_i=1}^{s_i} \hat{a}^{+\mu^i_{l_i}}_i|0\rangle, \ \ \{\hat{a}^i_{\mu^i},
\hat{a}_{\nu^j}^{j+}\}=-\eta_{\mu^i\nu^i}\delta^{ij}, \\
\label{lilijt} \hspace{-2ex} && \hspace{-2ex} \bigl( {l}_0,\,{l}^i,\, l^{ij},\,
t^{ij},\, g_0^i-(s_i-\textstyle\frac{d}{2})\bigr)|\Phi\rangle  = \bigl(\partial^\mu\partial_\mu,\,-i \hat{a}^i_\mu \partial^\mu,\,
\textstyle\frac{1}{2}\hat{a}^{i}_\mu \hat{a}^{j\mu}, \,\hat{a}^{i+}_\mu
\hat{a}^{j\mu},\, -\hat{a}^{i+}_\mu  \hat{a}^{\mu{}i }- s_i \bigr) |\Phi\rangle=\vec{0} .
\end{eqnarray}
 The set of operators $\{o_\alpha\}$
= $\bigl\{{{l}}_0, {l}^i, l^{ij}, t^{ij} \bigr\}$ (for $l_0$ corresponding to hamiltonian $H_0$ from BFV method \cite{BFV}), which generate  $(1+6)$ even and $3$ odd, ${l}^i$,
primary constraints  (\ref{lilijt}), and  number particles operators, $g_0^i$
  should form closed superalgebra with respect to $[\ ,\ \}$-multiplication and hermitian conjugation with inner product
  \begin{eqnarray}
\label{sproduct} \langle{\Psi}|\Phi\rangle & =  & \int
d^dx\sum_{s_1=0}^{[d/2]}\sum_{s_2=0}^{s_1}\sum_{s_3=0}^{s_{2}} \frac{(-1)^{\sum_ps_p}}{s_1!s_2! s_3!}\Psi^{\star}_{\mu^1[{s_1}],\mu^2[{s_2}], \mu^3[{s_3}]}(x)
\Phi^{\mu^1[{s_1}],\mu^2[{s_2}], \mu^3[{s_3}]}(x) .
\end{eqnarray}
 As the result, the set of $\{o_\alpha\}$ extended by means of the
operators,
\begin{eqnarray} \label{lilijt+} \hspace{-2ex} && \hspace{-2ex} \bigl({l}^{i+},\
l^{ij+},\ t^{ij+} \bigr)  = \bigl(-i \hat{a}^{i+}_\mu
\partial^\mu,\ \textstyle\frac{1}{2}\hat{a}^{j+}_\mu \hat{a}^{i\mu+},\
\hat{a}^{j+}_\mu \hat{a}^{i\mu}\bigr) ,\
\end{eqnarray}
forms superalgebra Lie of operators $\{o_I\}=$ $\{o_A;o_a,o_a^+; g_0^i\}$, for isometry subsuperalgebra of $\{o_A\} = $ $\{ {l}_0, {l}^{i}, {l}^{i+}\}$ and subalgebra $\{o_a,o_a^+; g_0^i\}$ with operators of second-class (holonomic) constraints $\{o_a,o_a^+\}$ for  $o_a=(l^{ij}, t^{ij})$.The multiplication law for  $o_I$ is given by the compact relations
\begin{equation}\label{geninalg}
    [o_I,\ o_J]= f^K_{IJ}o_K: \ [o_a,\; o_b^+\} = f^c_{ab} o_c +\Delta_{ab}(g_0^i), \  f^K_{IJ}= - (-1)^{\varepsilon(o_I)\varepsilon(o_J)} f^K_{JI},
\end{equation}
with invertible matrix $\|\Delta_{ab}(g_0^i)\|$ in $\mathcal{H}^f$ on the solutions for the system $(o_a,o_A)|\Phi\rangle=0$.
Explicitly, the  structure constants $f^K_{IJ}$ determined from the
multiplication table~\ref{table in}. \hspace{-1ex}
{\begin{table}
{{\footnotesize
\begin{center}
\begin{tabular}{||c||c|c|c|c|c|c|c|c|c|c|c|c||}
\hline\hline \rule{0cm}{0.4cm}
\hspace{-0.7em} $[\; \downarrow, \rightarrow ] $ \hspace{-0.7em} &
  $t^{\prime}_{12}$ & $t^{\prime}_{13}$ & $t^{\prime}_{23}$ &
 $t^{+\prime}_{12}$ & $t^{+\prime}_{13}$ & $t^{+\prime}_{23}$ &
$l^{+\prime}_{12}$ & $l^{+\prime}_{13}$ &$l^{+\prime}_{23}$ & $g^{\prime}_{01}$ & $g^{\prime}_{02}$ & $g^{\prime}_{03}$ \\
\hline \hline \rule{0cm}{0.4cm} $t^{\prime}_{12}$ & $ 0 $ & $ 0 $ & $ -t^{\prime}_{13} $&
\hspace{-0.7em} {\tiny{}$ g^{\prime}_{01} -g^{\prime}_{02}$}\hspace{-0.5em} & $ t^{+\prime}_{23} $ & $ 0 $ &  $ 0 $ & $ 0 $ & $ -l^{+\prime}_{13} $ &
 $ -t^{\prime}_{12} $ & $t^{\prime}_{12} $ & $ 0 $ \\
  \hline \rule{0cm}{0.4cm}  $t^{\prime}_{13}$ & $ 0 $ & $ 0 $ & $ 0 $ &
    $ t^{\prime}_{23} $ & \hspace{-0.5em}{\tiny{}$ g^{\prime}_{01} -g^{\prime}_{03} $}\hspace{-0.5em} &  $ -t^{\prime}_{12} $ & $ 0 $ & $ 0 $ & $ l^{+\prime}_{12} $ &
     $ -t^{\prime}_{13}$ & $ 0 $ & $ t^{\prime}_{13} $  \\
  \hline\rule{0cm}{0.4cm}  $t^{\prime}_{23}$ & $ t^{\prime}_{13} $ & $ 0  $ & $ 0 $& $ 0 $ &
  $ -t^{+\prime}_{12} $ &\hspace{-0.7em}{\tiny{} $ g^{\prime}_{02} -g^{\prime}_{03} $ }\hspace{-0.9em}&
  $ 0 $ & $ -l^{+\prime}_{12} $ & $ 0 $ & $ 0 $  & $- t^{\prime}_{23} $ & $ t^{\prime}_{23} $ \\
\hline \rule{0cm}{0.4cm} $t^{+\prime}_{12}$
    &\hspace{-0.7em}{\tiny{}  $-g^{\prime}_{01} +g^{\prime}_{02} $}\hspace{-0.5em} & $ -t^{\prime}_{23} $ & $0 $&
     $0$ & $0$ & $t^{+\prime}_{13}$ & $ 0 $ &
    $-l^{+\prime}_{23}$ &  $0$  & $t^{+\prime}_{12}$ & $ -t^{+\prime}_{12}$ & $ 0 $ \\
\hline  \rule{0cm}{0.4cm}  $t^{+\prime}_{13}$ &
 $- t^{+\prime}_{23} $ & \hspace{-0.7em}{\tiny{} $ -g^{\prime}_{01} +g^{\prime}_{03} $}\hspace{-0.5em} & $ t^{+\prime}_{12} $&
 $0 $ & $0$ &$0$ & $l^{+\prime}_{23}$ &
     $0$  & $ 0 $ & $t^{+\prime}_{13}$ & $0$  & $-t^{+\prime}_{13}$\\
\hline  \rule{0cm}{0.4cm}   $t^{+\prime}_{23}$ &
 $0$ & $t^{\prime}_{12}$ &\hspace{-0.7em} {\tiny{}$ -g^{\prime}_{02} +g^{\prime}_{03} $}\hspace{-0.9em}&
 $-t^{+\prime}_{13} $ & $ 0$ & $ 0 $ &
      $-l^{+\prime}_{13}$ & $0$ & $0$ &
     $0$ &  $t^{+\prime}_{23}$  & $ -t^{+\prime}_{23}$ \\
     \hline \rule{0cm}{0.5cm}  $l^{\prime}_{1}$
    &  $ -l^{\prime}_{2} $ & $-l^{\prime}_{3} $ & $ 0 $&
     $ 0 $ & $ 0 $ & $ 0 $&
     $ \frac{l^{+\prime}_{2}}{2} $ & $ \frac{l^{+\prime}_{3}}{2} $ & $0 $ &
      $ l^{\prime}_{1} $ & $ 0$ & $ 0 $ \\[0.1cm]
     \hline \rule{0cm}{0.5cm}  $l^{\prime}_{2}$ &  $ 0 $ & $ 0 $ & $-l^{\prime}_{3} $&
$ -l^{\prime}_{1}$ & $0 $ & $0 $& $ \frac{l^{+\prime}_{1}}{-2} $ & $0$&$0$ &
 $ 0 $ &  $ l^{\prime}_{2} $ & $ 0 $ \\[0.1cm]
      \hline \rule{0cm}{0.5cm}   $l^{\prime}_{3}$ &  $0 $ & $0 $ & $ 0$&
  $ 0$ & $-l^{\prime}_{1} $ & $-l^{\prime}_{2} $& $ 0 $ & $ \frac{l^{+\prime}_{1}}{-2} $ & $ \frac{l^{+\prime}_{2}}{-2} $ & $ 0 $ &
 $ 0 $ & $ l^{\prime}_{3} $  \\[0.1cm]
\hline\rule{0cm}{0.5cm}  $l^{+\prime}_{1}$
    &  $0 $ & $0 $ & $0 $&
     $ l^{+\prime}_{2} $ & $l^{+\prime}_{3} $ & $ 0$&
     $ 0$ & $0 $ & $0 $ &
      $ -l^{+\prime}_{1}  $ & $ 0$ & $ 0 $ \\
\hline \rule{0cm}{0.5cm}  $l^{+\prime}_{2}$ &  $l^{+\prime}_{1} $ & $0 $ & $0 $&
$0 $ & $ 0$ & $ l^{+\prime}_{3}$& $0 $ & $ 0$ & $ 0$ &
$0 $ &  $ -l^{+\prime}_{2}  $ & $ 0 $ \\
 \hline \rule{0cm}{0.5cm}  $l^{+\prime}_{3}$ &  $0 $ & $ l^{+\prime}_{1} $ & $l^{+\prime}_{2} $&   $0 $ & $ 0$ & $ 0$&
   $ 0$ & $ 0$ & $0$ &
   $0 $ & $0 $ & $ -l^{+\prime}_{3} $  \\
 \hline  \rule{0cm}{0.5cm}  $l^{\prime}_{12}$
   &  $ 0 $ & $ l^{\prime}_{23} $ & $ -l^{\prime}_{13} $&
   $ 0 $  & $ 0 $  & $ 0 $  &
   $\frac{g^{\prime}_{01}+g^{\prime}_{02}}{-4} $ &
   $\frac{t^{+\prime}_{23}}{4}$ &
    $\frac{t^{+\prime}_{13}}{-4}$ &
     $l^{\prime}_{12}$
     & $l^{\prime}_{12}$  & $0$  \\[0.1cm]
\hline \rule{0cm}{0.5cm}  $ l^{\prime}_{13}$ &
 $ -l^{\prime}_{23} $ & $ 0 $ & $ 0 $&
 $ 0 $ & $ 0 $ & $ -l^{\prime}_{12} $ &
 $ \frac{t^{\prime}_{23}}{4} $  &
 $\frac{g^{\prime}_{01}+g^{\prime}_{03}}{-4}$  & $ \frac{t^{+\prime}_{12}}{4} $ & $l^{\prime}_{13}$ & $ 0 $  & $l^{\prime}_{13}$  \\[0.1cm]
\hline \rule{0cm}{0.5cm} $ l^{\prime}_{23}$
    &  $0 $ & $ 0 $ & $ 0 $&
    $ -l^{\prime}_{13} $ & $ l^{\prime}_{12} $ & $ 0 $& $ -\frac{t^{\prime}_{13}}{4} $ & $ \frac{t^{\prime}_{12}}{4} $ &
    $\frac{g^{\prime}_{02}+g^{\prime}_{03}}{-4}$ &
    $ 0 $ & $ l^{\prime}_{23}$  & $ l^{\prime}_{23}$\\[0.1cm]
\hline \rule{0cm}{0.5cm}  $l^{+\prime}_{12}$
   &  $ 0$ & $0 $ & $0 $&
   $0  $  & $-l^{+\prime}_{23} $  & $ l^{+\prime}_{13}$  & $0$ & $0$ & $0$ &  $-l^{+\prime}_{12}$
     & $-l^{+\prime}_{12}$  & $0$  \\
\hline \rule{0cm}{0.5cm}  $ l^{+\prime}_{13}$ &
 $0 $ & $ 0$ & $l^{+\prime}_{12} $&
 $ l^{+\prime}_{23}$ & $ 0$ & $ 0$ & $ 0$  & $0$  & $ 0 $ &
$-l^{+\prime}_{13}$ & $ 0 $  & $-l^{+\prime}_{13}$  \\
\hline  \rule{0cm}{0.5cm}  $ l^{+\prime}_{23}$
    &  $l^{+\prime}_{13} $ & $ -l^{+\prime}_{12} $ & $ 0$&
    $ 0$ & $0 $ & $0 $&
     $ 0$ & $0 $ & $0$ &
    $ 0 $ & $ -l^{+\prime}_{23}$  & $ -l^{+\prime}_{23}$\\
\hline \rule{0cm}{0.5cm} $g^{\prime}_{01}$
    &  $t^{\prime}_{12} $ & $t^{\prime}_{13} $ & $ 0 $&
     $-t^{+\prime}_{12} $ & $-t^{+\prime}_{13} $ & $ 0$& $ l^{+\prime}_{12}$ & $ l^{+\prime}_{13}$ & $0 $ & $ 0 $ &
 $ 0$ & $ 0 $ \\
\hline \rule{0cm}{0.5cm}  $g^{\prime}_{02}$ &  $-t^{\prime}_{12} $ & $0 $ & $ t^{\prime}_{23}$&
$t^{+\prime}_{12} $ & $0 $ & $- t^{+\prime}_{23}$& $l^{+\prime}_{12} $ & $0 $ & $l^{+\prime}_{23} $ &
 $ 0$ &  $ 0 $ & $ 0 $ \\
 \hline \rule{0cm}{0.5cm}  $g^{\prime}_{03}$ &  $ 0$ & $-t^{\prime}_{13} $ & $-t^{\prime}_{23} $&
  $0 $ & $t^{+\prime}_{13} $ & $t^{+\prime}_{23} $& $0 $ & $ l^{+\prime}_{13}$ & $l^{+\prime}_{23} $ &
  $ 0$ &  $0 $ & $ 0 $  \\
\hline\hline
\end{tabular}
\end{center}}} \vspace{-2ex}\caption{HS symmetry  superalgebra  $\mathcal{A}(Y[3],\mathbb{R}^{1,d-1})$.\label{table in} }\end{table}
  Note the  indices $i,j,k$ are
raising and lowering by  Euclidian metric tensors
$\delta^{ij}$, $\delta_{ij}$, $\delta^{i}_{j}$}

The algebra of the operators $o_I$ is called the \emph{integer
HS symmetry superalgebra in Minkowski space with a Young
tableau having $3$ columns} and denoted  as $\mathcal{A}(Y[3],
\mathbb{R}^{1,d-1})$.

The subalgebra with holonomic constraints $(o_a,o_a^+)$ is isomorphic to algebra $so(3,3)$.
Its presence means for the construction of LF with complete BRST operator a necessity of conversion into a system of deformed first-class operator constraints $(o_a,o_a^+)\to(O_a,O_a^+)$.
Within additive conversion receipt, see for instance, \cite{1211.1273}, \cite{ashsf2},
\cite{BurdikPashnev},   which
implies the enlarging of $o_I$ to $O_I = o_I + o'_I$, where
additional parts $o'_I$ are given on a new Fock space
$\mathcal{H}'$ being independent on $\mathcal{H}^f$ in a sense that
$\mathcal{H}'\bigcap \mathcal{H}^f = \emptyset$. The operators $O_I,  o'_I$ satisfy the same algebra as one (\ref{geninalg}) for $o_I$. From Verma module construction for $so(3,3)$  with basis  $\{|\vec{N}\rangle_V\} $
\begin{eqnarray}
 &&|\vec{N}\rangle_V \ \stackrel{def}{=}\  |\vec{n}_{ij}, \vec{p}_{rs} \rangle_V \ = \ \prod_{i=1}^{2}\Bigr[\prod_{j=i+1}^{3}
\bigl(l^{\prime +}_{ij}\bigr){}^{n_{ij}}\Bigl]\prod_{r=1}^{2}\Bigr[\prod_{s=r+1}^{3}
\bigl(t^{\prime +}_{rs}\bigr){}^{p_{rs}}\Bigl] |0\rangle_V, \
  \label{module}\\
  &&  \texttt{for} \ \big(l^{\prime }_{ij}, t^{\prime}_{rs}; g_0^{i\prime}\big)|0\rangle_V = \big(\vec{0}_{ij}, \vec{0}_{rs}; h^{i}\big)|0\rangle_V, \ \ {}_V\langle0|0\rangle_V =1 \nonumber
\end{eqnarray}
(with some $\mathbb{R}$-valued numbers $h^{i}$)
 it follows from the mapping between basis $\{|\vec{N}\rangle_V\}$ (\ref{module})
 and the one $\{|\vec{N}\rangle\} \equiv$ $\{ \prod_{i=1}^{2}\prod_{j= i+1}^3\bigl(b^{+}_{ij}\bigr){}^{
 n_{ij}}\prod_{r=1}^{2}\prod_{s= r+1}^3\bigl(d^{+}_{rs}\bigr){}^{p_{rs}}|0\rangle\}$ in new Fock space
$\mathcal{H}'$, the scalar oscillator representation for $o'_I$
 \begin{eqnarray}
 g^{1\prime}_{0}  & = & b_{12}^{+} b_{12}+b_{13}^{+} b_{13}- d_{12}^{+} d_{12}- d_{13}^{+} d_{13}+h^1 \nonumber \\
 g^{2\prime}_{0} & = & b_{12}^{+} b_{12}+b_{23}^{+} b_{23}- d_{23}^{+} d_{23}+ d_{12}^{+} d_{12}+h^2 \label{g'03Faa}\\
 g^{3\prime}_{0}  & = &  b_{13}^{+} b_{13}+b_{23}^{+} b_{23}+ d_{13}^{+} d_{13}+ d_{23}^{+} d_{23}+h^3
 \,, \nonumber
 \\
 \big(l^{\prime+}_{ij},\,t^{+\prime}_{12},\,t^{+\prime}_{13},\, t^{+\prime}_{23}\big) & = &   \big(b_{ij}^+,\, d_{12}^{+}- b_{23}^{+}b_{13},\,d_{13}^{+}+b_{23}^{+}b_{12},\,
    d_{23}^{+}-d_{13}^{+}d_{12}-b_{13}^{+}b_{12}\big)
   \label{l'+ijFaa}
 \end{eqnarray}
 for the elements $l^{\prime }_{lm}$ for  $l<m$
\begin{eqnarray}
  l^{\prime}_{12} & = & -\frac{1}{4} \Bigl[\bigl(h^1+h^2 +\sum_{i>1}^3b_{1i}^+b_{1i} +b_{23}^+b_{23}-\sum_{i<3}^3d_{i3}^+d_{i3}\bigr)b_{12}-\bigl(d^+_{23}-d^+_{13}d_{12}\bigr)b_{13}+b_{23}d^+_{13}\Bigr]   \label{l'12Faa} \\
l^{\prime}_{13} & = & \frac{1}{4} \bigl(-h^1-h^3 -b_{12}^+b_{12}- b_{13}^+b_{13} -b_{23}^+b_{23}+d_{12}^+d_{12}- d_{23}^+d_{23} \bigr)b_{13}\label{l'13Faa} \\
&&  +\frac{1}{4}\bigl[\bigl(h^2-h^3 - d_{23}^+d_{23} \bigr)d_{23}-d_{12}^+d_{13}\bigr]b_{12}+\frac{1}{4} d^+_{12}b_{23} \nonumber\\
 l^{\prime}_{23} & = & -\frac{1}{4}\bigl(h^2+h^3 +b_{12}^+b_{12}+ b_{13}^+b_{13} +b_{23}^+b_{23}+d_{12}^+d_{12}+ d_{13}^+d_{13} \bigr)b_{23}+\frac{1}{4}b_{13}d_{13}d^+_{23}\label{l'23Faa} \\
&& -\frac{1}{4}\bigl(h^1-h^3 -d_{12}^+d_{12}-d_{13}^+d_{13}- d_{23}^+d_{23} \bigr)b_{12}d_{13}\nonumber\\
&&  +\frac{1}{4}\bigl(h^1-h^2 -d_{12}^+d_{12}-d_{13}^+d_{13}+ d_{23}^+d_{23} \bigr)b_{13}d_{12}-\frac{1}{4}\bigl(h^2-h^3 - d_{23}^+d_{23} \bigr)b_{12}d_{12}d_{23}\nonumber
 \end{eqnarray}
 and for  $t^{\prime }_{lm}$,
 \begin{eqnarray}
 t^{\prime}_{12}  & = & -b_{13}^{+} b_{23} + \bigl(h^1-h^2 - d_{13}^+d_{13}+ d_{23}^+d_{23}- d_{12}^+d_{12}\bigr)d_{12} + d_{23}^+d_{13},
 \label{t'12Faa} \\
 t^{\prime}_{13}  & = & b_{12}^{+} b_{23} + \bigl(h^1-h^3- d_{12}^+d_{12} - d_{13}^+d_{13}- d_{23}^+d_{23}\bigr)d_{13} + \bigl(h^2-h^3 - d_{23}^+d_{23}\bigr)d_{12}d_{23}, \label{t'13Faa}  \\
 t^{\prime}_{23}  & = &  -b_{12}^+b_{13} + \bigl(h^2-h^3 - d_{23}^+d_{23}\bigr)d_{23} - d_{12}^+d_{13}. \label{t'23Faa}
\end{eqnarray}
The operators $o'_I$ are hermitian with respect to modified (by operator  $K'$
intertwining $\{|\vec{N}\rangle_V\}$ and $\{|\vec{N}\rangle\}$ (\ref{fin K})) scalar product. Having in mind the rules of the dimensional reduction
from  massless theory in $\mathbb{R}^{1,d}$  space-time to massive one in $\mathbb{R}^{1,d-1}$ with fermionic oscillators $f_i^+,f_i$,
\begin{align} \label{reduction}
   &\partial^{M} = (\partial^{\mu}, -\imath m)\,, &&\hat{a}^{M}_i = (\hat{a}^{\mu}_i, f_i)\,, &&
   \hat{a}^{M{}+}_i = (\hat{a}^{\mu{}+}_i, f_i^+)\,,  \\
   &M=0,1,\ldots ,d\,, && \mu=0,1,\ldots ,d-1\,, && \eta^{MN} =
   diag (1,-1,\ldots,-1,-1)\,,\label{reduction1}
\end{align}
the converted operators $O_I$ now for massive case are derived as well.

\subsection{\label{sec1:level2}BRST operator and Lagrangian formulation}

Complete BRST operator $Q$ is included into nilpotent operator $Q'$, (due to number particle operators $G_0^i$ presence in $O_I$) and may be easily found according to prescription (\ref{QQc}).
 $Q'$ can be presented via incomplete  BRST operator   $Q_c$ with only differential first-class constraints $\{l_0,l_i,  l_i^+\}$ and with \emph{complete BRST-extended} converted second-class  constraints $\{\mathcal{O}_a. \mathcal{O}^+_a\}$ as
\begin{eqnarray}
Q'& = &   Q_c +  \sum_{i<j}\Big(\eta_{ij}\mathcal{L}{}^+_{ij} + \mathcal{L}_{ij}\eta^+_{ij}+  \mathcal{T}_{ij}\vartheta_{ij}^+ +\vartheta_{ij}\mathcal{T}^+_{ij}
\Big)+ \sum_i[\eta^G_{i}\sigma^i+ \imath \mathcal{B}^i\mathcal{P}_{i}^G ] , \label{Q'Q}
\end{eqnarray}
with definite operators $\mathcal{B}^i$, generalized spin operator $\sigma^i$ and with complete traceless and Young constraints (and theirs  hermitian conjugated ones)
\begin{eqnarray}
\sigma^i\hspace{-0.2em}&=&\hspace{-0.2em} G_{0i}- q_i^+ p_i  - q_i p_i^+  +  \sum_{j\ne i}[\eta_{ij}^+ \mathcal{P}_{ij} - \eta_{ij} \mathcal{P}^+_{ij}] - \sum_{j>i}[\vartheta^+_{ij} \lambda_{ij} -
\vartheta_{ij}\lambda^+_{ik}]+
  \sum_{j<i} [\vartheta^+_{ji}
\lambda_{ji} - \vartheta_{ji}\lambda^+_{ji}],  \label{gspin}\\
\mathcal{L}_{ij}\hspace{-0.2em}&=&\hspace{-0.2em} \widehat{L}'_{ij}  +\hspace{-0.15em}\sum_{p<j}\hspace{-0.2em}\vartheta_{ip}^+\mathcal{P}_{pj}
+\hspace{-0.2em}\sum_{j<p\leq 3}\hspace{-0.3em} \vartheta_{jp}^+\mathcal{P}_{ip} -\hspace{-0.2em}\sum_{j<p\leq 3}\hspace{-0.3em}\vartheta_{ip}^+\mathcal{P}_{j p}+\hspace{-0.1em}\frac{1}{4}\Big\{\hspace{-0.1em}\sum_{3\geq  p>j}\hspace{-0.3em}\big[\eta_{ip}\lambda^+_{jp}-\hspace{-0.1em}\eta_{jp}\lambda^+_{ip}\big]  + \sum_{p<j}\hspace{-0.2em}\eta_{pj}\lambda^+_{ip}
    \Big\} ,\label{BRSTextTr}\\
 \mathcal{T}_{ij} &=&\widehat{T}'_{ij}+ \sum_{p>j} \vartheta^+_{jp}\lambda_{ip} - \sum_{3\geq p<j}[\vartheta_{ip}\lambda_{pj}-\vartheta_{pj}\lambda_{ip}]
    +\sum_{3\geq p>j}\eta_{jp}\mathcal{P}_{ip}^+ -  \sum_{i<p<j} \eta_{pj}\mathcal{P}_{ip}^+ + \sum_{p<i}\eta_{pj}
 \mathcal{P}_{pi}^+,\label{BRSTextYo}\\
    && \texttt{for} \  \big(\widehat{L}'_{ij},  \widehat{T}'_{ij}\big) \ = \ \big(L_{ij} + \frac{1}{2} q_{[i} p_{j]},\,{T}_{ij}+(q_jp_i^+ +q_i^+p_j)\big)\label{incompleteconstr}
\end{eqnarray}
to be \emph{incomplete BRST-extended} operator constraints $\big(\widehat{L}'_{ij},  \widehat{T}'_{ij}\big)$, Sum of   $Q_c$ with rest term
  without  number particles ghost coordinates, $\eta^G_{i}$, and momenta, $\mathcal{P}_{i}^G$ determines complete BRST operator $Q$ for only converted
constraint system with extracted zero-mode ghosts
\begin{eqnarray}
Q& =&  Q_c + \sum_{i<j\leq 3}\big( \mathcal{L}_{ij}\eta^+_{ij}+  \mathcal{T}_{ij}\vartheta_{ij}^+ +h.c,\big), \ \  (\epsilon, gh_H)Q=(1,1).  \label{separation1}
\end{eqnarray}
(for $gh_H(\mathcal{C}^I) = - gh_H(\mathcal{P}_I)=1$). Here, the Grassman-even ghost coordinates  $q_i^+$, $q_i$  momenta  $p_j$, $p_j^{+}$ for Grassman-odd basis elements $l_i$,
$l_i^+$  of the superalgebra $\mathcal{A}_C(Y[3],\mathbb{R}^{1,d-1})$,   respectively, and Grassman-odd ghost coordinates $\eta_0$,  $\eta_{ij}^+$, $\eta_{ij}$, $\vartheta_{rs}^+$, $\vartheta_{rs}$, $\eta_{i}^G$
and  momenta $\mathcal{P}_0$,  $\mathcal{P}_{ij}$, $\mathcal{P}_{ij}^+$, $\lambda_{rs}$, $\lambda_{rs}^+$, $P^G_{i}$ for Grassman-even basis elements $l_0$,  $L_{ij}$,
$L_{ij}^+$, $T_{rs}$,  $T_{rs}^+$,  $G_0^{i}$ of $\mathcal{A}_C(Y[3],\mathbb{R}^{1,d-1})$,  respectively, are introduced subject to non-trivial commutation relations,
\begin{eqnarray}
 && [q_i^+,p_j]=[p_i^+,q_j]=\delta_{ij},
 \qquad \{\eta_0, \mathcal{P}_0\}=\imath, \qquad \{\eta^G_{i},\mathcal{P}^G_{j}\}=\imath \delta_{ij}, \\
&&\{\eta_{ij}^+,\mathcal{P}_{i'j'}\}=\{\eta_{ij},\mathcal{P}_{i'j'}^+\}=\delta_{ij,i'j'}
\{\vartheta_{rs},\lambda_{r's'}^+\}=\{\vartheta_{rs}^+,\lambda_{r's'}\}=\delta_{rs,r's'}.
\label{ghost_def}
\end{eqnarray}
The hermitian conjugation is understood in the sense of the rule
 \begin{eqnarray}\label{HermQ}
Q^{\prime +}K\  =\ K Q',  \ (\mathcal{L}_{ij}, \mathcal{T}_{ij})^+K\  =\ K(\mathcal{L}^+_{ij}, \mathcal{T}^+_{ij})  \texttt{   with    } K\ = \ 1 \otimes K' \otimes 1_{gh}.
\end{eqnarray}
with non-degenerate Grassmann-even operator $K$  to be determined on the total Hilbert space, $\mathcal{H}_{tot} = $  $\mathcal{H}^f \otimes \mathcal{H}'$ $\otimes \mathcal{H}_{gh}$.  It  can be constructed
from the whole set of oscillators, whereas $K'$  from only  $b_{ij}^{(+)}$ and $d_{rs}^{(+)}$:
\begin{eqnarray}
 K'&=&\hspace{-0.3em}\sum_{n_{ij}, p_{rs},n'_{ij}, n'_{rs} = 0}^{\infty}
\prod_{I,r=1}^2\prod_{j\geq i,s\geq r}^3 \hspace{-0.2em}\frac{{b}_{1j}^{+n_{ij}}}{(n_{ij})!}\frac{{d}_{rs}^{+p_{rs}}}{(p_{rs})!}|0\rangle{}_V\langle \vec{n}_{ij},\vec{p}_{rs}
|\vec{n}{}'_{ij},\vec{p}{}'_{rs}\rangle_V
  \langle
0|\hspace{-0.2em}\prod_{I,r=1}^2\prod_{j\geq i,s\geq r}^3 \hspace{-0.3em} \frac{{b}_{1j}^{n'_{ij}}}{(n'_{ij})!}\frac{{d}_{rs}^{p'_{rs}}}{(p'_{rs})!},
 \label{fin K}
\end{eqnarray}
Thus, we have  constructed a Hermitian complete BRST operator for the HS symmetry superalgebra $\mathcal{A}_C(Y[3],\mathbb{R}^{1,d-1})$ of converted operators $O_I$.
Including oscillators $f_i^+$ and the ghost operators, we extend our basic vector  $|\Phi\rangle$ (\ref{PhysState}) given in $\mathcal{H}^f$  to
\begin{eqnarray}
|\chi\rangle & = &
\sum_{\{n\}_b=0}^{\infty}
\sum_{\{n\}_f=0}^{1}
\eta_{0}^{n_{\eta_{0}}}
\prod_{l=1}^{3}
(\eta_{l}^{G})^{ n_{l}}
q_l^{+n_{q_l}}
p_l^{+n_{p_l}}f_l^{+n_{f_l}}\prod_{1\leq i<j\leq 3}\eta_{ij}^{+n_{\eta_{ij}}}\mathcal{P}_{ij}^{+n_{P_{ij}}}{b}_{1j}^{+n_{ij}}
\label{extState}
\\
& \times &
\prod_{1\leq r<s\leq 3}\vartheta_{rs}^{+n_{\vartheta_{rs}}}
\lambda_{rs}^{+n_{\lambda_{rs}}}
{d}_{rs}^{+p_{rs}}
\left|\chi_{n_{\eta_{0}}n_{\eta_{ij}}n_{\vartheta_{rs}}n_{P_{ij}}n_{\lambda_{rs}}n_{l} n_{q_l}n_{p_l}n_{f_l} n_{ij}p_{rs}}(\hat{a}{}_i^+)\rangle\right. ,
 \nonumber
\end{eqnarray}
where
$\{n\}_b = n_{q_l},n_{p_l},n_{ij}, p_{rs}$ and $\{n\}_f = n_{\eta_0},n_{\eta_{ij}},n_{P_{ij}},n_{\vartheta_{rs}},n_{\lambda_{rs}}, n_{f_l}$

From the  BRST-like complex, with equation determining the
physical vector, $Q'|\chi\rangle$ = $0$, (for $|\chi\rangle =
|\chi^{0}\rangle$) and from the set of reducible gauge
transformations, $\delta|\chi\rangle$ = $Q' |\chi^{1}\rangle$,
 $\ldots$,
$\delta|\chi^{s-1}\rangle = Q'|\chi^{s}\rangle$, for $gh(|\chi^{n}\rangle)=-n $, $n=0,...,s$, for some integer $s>0$
in the representation in $\mathcal{H}_{gh}$   in accordance with representation (\ref{extState}),
\begin{eqnarray}
 \left( q_i,  p_i,  \eta_{12},  \mathcal{P}_{12}, \vartheta_{12}, \lambda_{12}, \mathcal{P}_0\right) | 0 \rangle &=&0, \; i = 1, 2 \label{ghreps}
\end{eqnarray}
it follows a finite
sequence of relations underlying the $\eta^i_G$-independence of
all of the above homogeneous in ghost number vectors:
\begin{align}
\label{Qchi} & Q|\chi\rangle=0, && (\sigma^i+h_i)|\chi\rangle=0,
&& \left(\varepsilon, {gh}_H\right)(|\chi\rangle)=(\varepsilon_\chi,0),
\\
& \delta|\chi\rangle=Q|\chi^{1}\rangle, &&
(\sigma^i+h_i)|\chi^{1}\rangle=0, && \left(\varepsilon,
{gh}_H\right)(|\chi^{1}\rangle)=(\varepsilon_\chi+1,-1), \label{QLambda}
\\
& \ldots\ldots && \ldots\ldots && \ldots\ldots \nonumber \\
& \delta|\chi^{n-1}\rangle=Q|\chi^{n}\rangle, &&
(\sigma^i+h_i)|\chi^{n}\rangle=0, && \left(\varepsilon,
{gh}_H\right)(|\chi^{n}\rangle)= (\varepsilon_\chi +n \hspace{-0.5em}\mod{}2,-n). \label{QLambdai}
\end{align}
By one of the peculiarity of BRST approach to mixed-antisymmetric HS fields is the fact that the Grassmann parity of any of $|\chi^n\rangle$  depends on the values of spin.
The solutions of middle set of equations in
(\ref{Qchi})--(\ref{QLambdai}) determine values of spin for the eigenvectors $\{|\chi^{n}\rangle_{{s[3]}}\}$ of  generalized spin operator
$(\sigma^1, \sigma^2, \sigma^3)$ and  possible values for
 parameters $h_i$:
\begin{eqnarray}
(h_i^{s[3]}, h_i^{m|s[3]}) & = & \left( - s_i + \textstyle \frac{d-6}{2} + 2i \right)(1,1) +\left(0, {1}/{2}\right), \ \ i=1,2,3 .
\label{h}
\end{eqnarray}
As the result the operator $Q$ is nilpotent on the solutions for the middle set and  the  first equation in (\ref{Qchi}): $Q|\chi\rangle_{{s[3]}}=0$ is Lagrangian EoM and follows from gauge-invariant LF
\begin{eqnarray}
\mathcal{S}_{s[3]} = \int\hspace{-0.2em}d\eta_0\hspace{-0.1em}  {}_{s[3]}\langle \chi
|K_{s[3]} Q_{s[3]}| \chi \rangle_{s[3]} \hspace{-0.1em}\sim   \hspace{-0.1em}\int \hspace{-0.2em} d^dx \hspace{-0.15em}\Big(\hspace{-0.15em}\Phi_{\mu[{s_1}],\nu[{s_2}], \rho[{s_3}]}\hspace{-0.1em}(\partial^2\hspace{-0.1em}+\hspace{-0.1em}m^2)
\hspace{-0.1em}\Phi^{\mu[{s_1}],\nu[{s_2}], \rho[{s_3}]} + \hspace{-0.1em}\texttt{more}\Big),  \label{Scl}
\end{eqnarray}
(for substitution $(K_{s[3]}, Q_{s[3]})= (K, Q)|_{h^i=h^i(s[3])}$) with reducible gauge transformations
\begin{align}
\label{dx0} &\delta|\chi^0 \rangle_{s[3]}
=Q_{s[3]}|\chi^1\rangle_{s[3]}\,, \ldots,
\delta|\chi^{l} \rangle_{s[3]}
=Q_{s[3]}|\chi^{l+1}\rangle_{s[3]} ,\ l=\sum_{i=1}^3 s_i+2\,,
\end{align}
for independent gauge parameter $|\chi^{l+1} \rangle$.
Here the usual inner product for the creation and
annihilation operators is assumed with measure $d^dx$ over
Minkowski space with additional terms: "\emph{more}" with auxiliary fields differed for massless and massive basic HS field $\Phi^{\mu[{s_1}],\nu[{s_2}], \rho[{s_3}]}(x)$.

The LF appears by gauge theory for the particle with $(m, s[3])$
 of $(\sum_{i=1}^3 s_i+2)$ stage reducibility.

\section{\label{sec2:level1}Lagrangian formulation with incomplete BRST operator}

In this case the configuration space of fields contains less auxiliary fields as compared to one for LF with complete  BRST operator.
The LF is characterized by incomplete nilpotent, $Q_c$, BRST operator, operator of generalized spin $\sigma^i_c$ and BRST-extended operator holonomic constraints $\big(\widehat{L}_{ij},  \widehat{T}_{ij}\big)$ from (\ref{incompleteconstr})    not depending on conversion oscillators $b^{(+)}_{ij}, d^{(+)}_{rs}$
\begin{equation}\label{holconstr}
 \sigma^i_c = \sigma^i_c = g_{0i}- q_i^+ p_i  - q_i p_i^+,\quad
 \big(\widehat{L}_{ij},  \widehat{T}_{ij}\big) =  \big(\widehat{L}'_{ij},  \widehat{T}'_{ij}\big)\big|_{(b^{(+)}_{ij}, d^{(+)}_{rs})=0}
\end{equation}
The set of $Q_c, \sigma^i_c, \widehat{L}_{ij},  \widehat{T}_{ij}$ acts in Hilbert space $\mathcal{H}^f\otimes H_{gh}^{o_A}$ and  forms closed superalgebra. It
provides the consistent BRST $Q_c$-complex resolution to get LF with holonomic operator constraints imposed on whole set of restricted field $| \chi_c\rangle_{s[3]}$ and reducible gauge  parameter $| \chi_c^n \rangle_{s[3]}$ vectors. The latter vectors have the decomposition in oscillators like (\ref{extState}) but without conversion $b^{(+)}_{ij}, d^{(+)}_{rs}$ and ghost oscillators $ \eta_{ij}^{+}, \mathcal{P}_{ij}^{+}, \vartheta_{rs}^{+},
\lambda_{rs}^{+}$ for second-class constraint system.

The resulting LF for the particle with $(m, s[3])$ (using oscillators $f_i, f_i^+$ for $m \ne 0$) after repeating
the same steps as in the previous section has the form
\begin{eqnarray}
&& \mathcal{S}_{c|s[3]}(\chi_c) = \int d \eta_0 \; {}_{s[3]}\langle\chi_c
|Q_c| \chi_c \rangle_{s[3]} \sim   \hspace{-0.1em}\int \hspace{-0.2em} d^dx \hspace{-0.15em}\Big(\hspace{-0.15em}\Phi_{\mu[{s_1}],\nu[{s_2}], \rho[{s_3}]}\hspace{-0.1em}(\partial^2\hspace{-0.1em}+\hspace{-0.1em}m^2)
\hspace{-0.1em}\Phi^{\mu[{s_1}],\nu[{s_2}], \rho[{s_3}]} + \hspace{-0.1em}\texttt{more}\Big)\  , \label{Sclred}\\
&&\Big(\delta; \  \widehat{L}_{ij}, \   \widehat{T}_{rs} \Big)| \chi^l_c \rangle_{s[3]} =
\Big(Q_c| \chi^{l+1}_c \rangle_{s[3]}; \ {0},\ 0\Big), \ l=0,1,...,L=\sum_is_i \label{Sr}
\end{eqnarray}
(with $\delta| \chi^{L+1}_c \rangle_{s[3]} = 0$). For $\sum_is_i=0$ (for scalar field) the theory appears by non-gauge one, whereas in general, it is the Abelian gauge model of $(\sum_is_i-1)$-th stage reducibility.

In case of $\hat{s}_3=0$ LFs with (in)complete BRST operator corresponds to known \cite{resh1} results for LFs for massless  (massive) tensor fields $\Phi_{\mu^1[{\hat{s}_1}],\mu^2[{\hat{s}_2}]}$ subject to $Y[\hat{s}_1,\hat{s}_2]$.

Note, the tensor expressions for the Lagrangian and tower of gauge transformations are easily  obtained in calculating the inner products to reduce the  oscillator presence.  Second, the component form for the LF with incomplete BRST operator without
off-shell holonomic constraints reveals the generalized gauge-invariant triplet  formulation for reducible $ISO(1,d-1)$ group massless and massive representations with many spins $s'[3]$ being fewer than $s[3]$.
Third, according to the general  result on equivalence for the dynamics for LFs with complete and incomplete BRST operator for irreducible  mixed-symmetric HS fields $\Psi_{\mu^1({{s}_1}),\mu^2({{s}_2}), \mu^3({{s}_3})}$  subject to 3-row Young tableaux  \cite{resh2} the same principal statement follows for\emph{ equivalence of  dynamics} generated by LFs  with actions $\mathcal{S}_{s[3]}(\chi)$ and $\mathcal{S}_{c|s[3]}(\chi_c)$ for the same field $\Phi_{\mu^1{[s_1]},\mu^2[{s_2}],\mu^3[{s_3}]}$ as well as with one determining by set of irrep conditions (\ref{Eq-0b})--(\ref{Eq-3b}).

\section{\label{sec3:level1} Deformation procedure for interacting vertices}

To this end   we introduce $p$, $p\geq 3$, copies of LFs (in adapting the model for Yang-Mills type interactions with gauge group $SU(N)$, for $p=N^2-1$) with vectors
$|\chi^{(t)}_c\rangle_{{s}[3]_t}$, reducible gauge parameters
$|\Lambda^{(t)n}_c\rangle_{{s}[3]_t}\equiv |\chi^{(t)n+1}_c\rangle_{{s}[3]_t}$,
respective vacuum vectors $|0\rangle^t$ and oscillators for
$t=1,...,p$ (with notation $(s[3])_t \equiv  (s[3]_1, ..., s[3]_t)$ for arbitrary sets of indices for different copies). It permits one to  determine the deformed  action, gauge transformations up to $r$-tic vertices, $r=3,4,...,e$  in  powers of  deformation  constant $g$, beginning from sum of $p$ LFs copies   for free   HS fields, then for cubic, quartic and so on vertices:
\vspace{-1ex}
\begin{eqnarray}\label{S[e]}
  && S^{(m)_p}_{[e]|(s[3])_p}[(\chi_c)_p] \  = \  \sum\nolimits_{t=1}^{p} \mathcal{S}_{c|s[3]_{t}}[\chi^{(t)}_c]   + \sum\nolimits_{o=1}^e g^o S^{(m)_p}_{o|(s[3])_p}[(\chi_c)_p],
  \end{eqnarray}
\vspace{-1ex}  where
\begin{eqnarray}\label{S[3]}
\hspace{-0.7em}&\hspace{-0.7em}&\hspace{-0.7em}  S^{(m)_p}_{1|(s[3])_p}[(\chi_c)_p] =   \sum_{1\leq i_1<i_2<i_3\leq p} \hspace{-1.0em} \int \prod_{j=1}^3 d\eta^{(i_j)}_0  \Big( {}_{s[3]_{i_j}}\langle \chi^{(i_j)}_c
  \big|  V^{(3)}_c\rangle^{(m)_{(i)_3}}_{(s[3])_{(i)_3}}+h.c. \Big)  , \\
  \label{S[4]}
\hspace{-0.9em}&\hspace{-0.9em}&\hspace{-0.9em}  S^{(m)_p}_{2|(s[3])_p}[(\chi_c)_p] =   \sum_{1\leq i_1<i_2<i_3<i_4\leq p}\hspace{-1.0em}  \int \prod_{j=1}^4 d\eta^{(i_j)}_0  \Big( {}_{s[3]_{i_j}}\langle \chi^{(i_j)}_c
  \big|  V^{(4)}_c\rangle^{(m)_{(i)_4}}_{(s[3])_{(i)_4}}+h.c. \Big)  , \\
   && \ldots \ \ldots\ \ldots \ \ldots\ \ldots \ \ldots\ \ldots \ \ldots\ \ldots \ \ldots \ \ldots \ \ldots\ \ldots \ \ldots \nonumber\\
    \label{S[+]}
\hspace{-0.9em}&\hspace{-0.9em}&\hspace{-0.9em}  S^{(m)_p}_{e|(s[3])_p}[(\chi_c)_p] =   \sum_{1\leq i_1<i_2<...<i_e\leq p}  \hspace{-1.0em}\int \prod_{j=1}^e d\eta^{(i_j)}_0  \Big( {}_{s[3]_{i_j}}\langle \chi^{(i_j)}_c
  \big|  V^{(e)}_c\rangle^{(m)_{(i)_e}}_{(s[3])_{(i)_e}}+h.c. \Big) ,
\end{eqnarray}
also for deformed $l$-th level gauge transformations: $\delta^l_{[e]}  |\Lambda^{(t)l}_c\rangle_{s[3]_{t}}=(\delta^l_0+\sum_{q=1}^eg^q\delta^l_q )|\Lambda^{(t)l}_c\rangle_{s[3]_{t}} $ :
\begin{eqnarray}\label{gt1}
\hspace{-1.1em}&\hspace{-1.1em}&\hspace{-1.1em} \delta^l_1|\Lambda^{(t)l}_c\rangle_{s[3]_{t}}\hspace{-0.1em} = \hspace{-0.1em} - \hspace{-1.0em}\sum_{1\leq i_1<i_2\leq p}\hspace{-0.5em} \int \hspace{-0.3em}\prod_{j=1}^2 d\eta^{(i_j)}_0 \hspace{-0.15em} \Big[\hspace{-0.1em} {}_{s[3]_{\{i_1}}\hspace{-0.25em}\langle \chi^{(\{i_1)}_c
  \big|  {}_{{s}[3]_{i_2\}}}\hspace{-0.25em}\langle \Lambda^{(i_2\})l+1}_c
  \big| {V}{}^{(3)l}_c\rangle^{(m)_{(i)_2t}}_{({s}[3])_{(i)_2{}t}}\hspace{-0.1em}+h.c. \hspace{-0.15em}\Big]\hspace{-0.1em}  ,
  \end{eqnarray}
 \begin{eqnarray}
   &\hspace{-0.5em}&\hspace{-0.5em} \ldots \ \ldots\ \ldots \ \ldots\ \ldots \ \ldots\ \ldots \ \ldots\ \ldots \ \ldots \ \ldots \ \ldots\ \ldots \ \ldots \nonumber\\
    \label{gte}
&\hspace{-0.5em}&\hspace{-0.5em}  \delta^l_e|\Lambda^{(t)l}_c\rangle_{{s}[3]_{t}} =   -\sum_{1\leq i_1<...<i_{e-1}\leq p} \hspace{-1.0em} \int \prod_{j=1}^{e-1} d\eta^{(i_j)}_0  \Big[{}_{{s}[3]_{{i_1}}}\langle \chi^{(\{i_1)}_c
  \big| \ldots {}_{{s}[3]_{{i_{e-2}}}}\langle \chi^{(i_{e-2})}_c
  \big| \\
  &\hspace{-0.5em}&\hspace{-0.5em} \phantom{\delta_2|\chi^{(j)}_c\rangle_{s_j}\ \ }\otimes
   {}_{{s}[3]_{i_{e-1}}}\langle\Lambda^{(i_{e-1}\})l+1}_c
  \big|  {V}{}^{(e)l}_c\rangle^{(m)_{(i)_{e-1}t}}_{({s}[3])_{(i)_{e-1}t}}+h.c. \Big] . \nonumber
\end{eqnarray}
Here we have used the notations $(\chi_c)_p= (\chi^{(1)}_c, \chi^{(2)}_c, ..., \chi^{(p)}_c)$, $|\Lambda^{(t)-1}_c\rangle \equiv |\chi^{(t)}_c\rangle$, the supersymmetrization  of indices $\{i_1,...,i_{e-1}\}$  with $r$-tic vertices $\big|  {V}{}^{(r)l}_c\rangle$, for $l=-1,0,...,L_t$.
The preservation for the interacting theory constructed from  initial  actions $\mathcal{S}^{m_t}_{0|\vec{s}_{k_t}}$, $t=1,...,p$  the number $N_t$  of physical degrees of freedom (p.d.f.)  determined by LFs for free HS field  with spin ${s}[3]_{t}$, requires  that the sum of all p.d.f. would be unchangeable, i.e. $\sum_{t}N_t= \mathrm{const}$. The property will be guaranteed, first, if the deformed  ac\-tion $S^{(m)_p}_{[e]|({s}[3])_{p}}$ will obey to sequence of new Noether identities in powers of $g$:
\begin{eqnarray}
       &g^1:& \delta_0 S^{(m)_p}_{1|(s[3])_p}[(\chi_c)_p]   + \delta_1 \mathcal{S}_{[0]|(s[3])_p}[(\chi_c)_p]  =0   , \label{g1} \\
    &g^2:& \delta_0 S^{(m)_p}_{2|(s[3])_p}[(\chi_c)_p]+ \delta_1 S^{(m)_p}_{1|(s[3])_p}[(\chi_c)_p] + \delta_2 \mathcal{S}_{0|(s[3])_p}[(\chi_c)_p] = 0 , \label{g2} \\
&& \ldots \ \ldots \  \ldots \  \ldots \ \ldots \ \ldots \ \ldots \ \ldots \ \ldots \ \ldots \ \nonumber \\
    &g^{e}:& \sum_{j=0}^e \delta_j S^{(m)_p}_{e-j|(s[3])_p}[(\chi_c)_p] =0 , \label{ge}
 \end{eqnarray}
 second, from $\delta^{l}_{[\infty]}\delta^{l-1}_{[\infty]}|\Lambda^{(t)l-1}_c\rangle_{s[3]_{t}}\vert_{\partial S^{(m)_p}_{[\infty]|(s[3])_p}=0} =0 $ for all  levels of
 gauge transformations:
 \begin{eqnarray}
{g^1}: && \Big(\delta^{l}_1\delta^{l-1}_0+ \delta^{{l}}_0\delta^{l-1}_1 \Big)|\Lambda^{(t)l-1}_c\rangle_{\vec{s}_{k_t}}\vert_{\partial S^{(m)_p}_{[1]}=0} =0 , \label{g0Ltttt1}  \\
{g^2}: && \Big(\delta^{l}_2\delta^{l-1}_0+ \delta^{{l}}_1\delta^{l-1}_1+\delta^{l}_0\delta^{l-1}_0 \Big)|\Lambda^{(t)l-1}_c\rangle_{s[3]_t} \vert_{\partial S^{(m)_p}_{[2]}=0} =0,\label{g0Ltttt2}\\
\vspace{-0.5ex}&& \ldots\ldots\ldots\ldots\ldots\ldots\ldots\ldots\ldots \nonumber \\
\vspace{-0.5ex} {g^e}:  &&  \Big(\delta^{l}_e\delta^{l-1}_0 + \textstyle\sum_{p=1}^e\delta^l_{e-p}\delta^{l-1}_p  \Big)|\Lambda^{(t)l-1}_c\rangle_{{s}[3]_{t}} \vert_{\partial S^{(m)_p}_{[e]}=0} =0 \label{g0Ltttte}
\end{eqnarray}
(for $\delta^{-1}\equiv \delta$  and $l=0,...,L-1$, where  $L= {\max}_{t} L_{(t)}$).
Third, we should take into account for influence of traceless $L^{(t)}_{ij}$ and Young-symmetry $T^{(t)}_{rs}$ constraints on the structure of vertices $\big|  V^{(q)}_c\rangle^{(m)_{(i)_q}}_{({s}[3])_{(i)_q}}$, $\big|  {V}{}^{(q)l}_c\rangle^{(m)_{(i)_q}}_{({s}[3])_{(i)_q}}$ for $q=3,4,...,e$.

\begin{figure}[h]
{\footnotesize\begin{picture}(9,3)
\put(-35.5, -27.5){\includegraphics[scale=0.30]{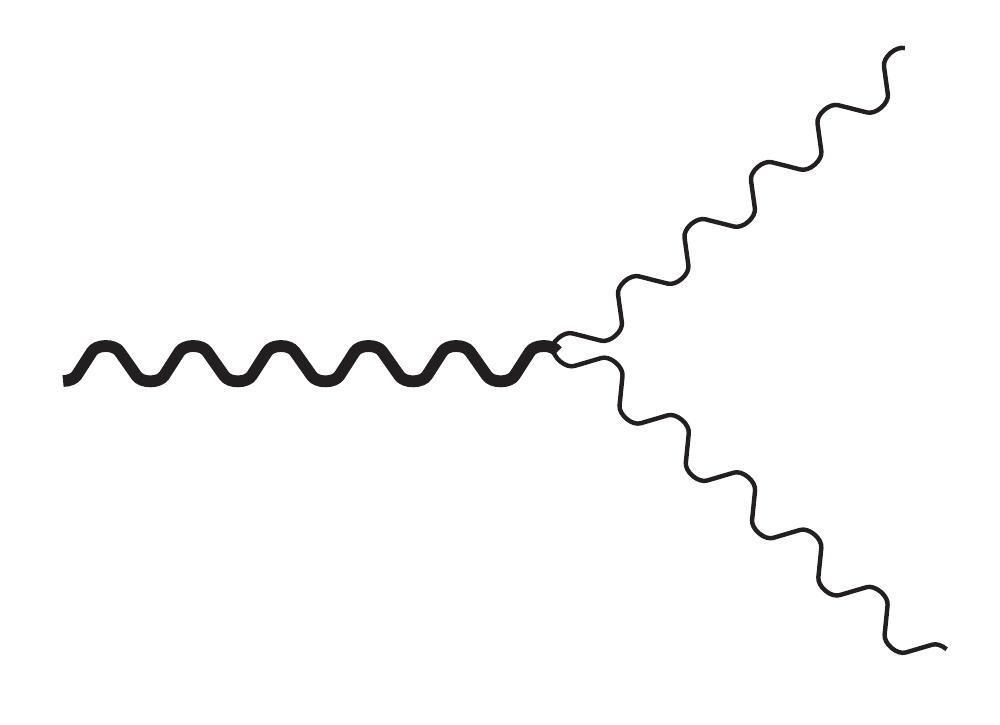}}
\put(-140.5, 20.5){$ |{V}{}^{(3)}\rangle^{(0,0,m)}_{(\lambda[3]_1 ,\lambda[3]_2,  s[3]_3)}\  \equiv $\  }
\put(-20.5, 35.5){$ \big(m, s[3]_3\big)$ }
\put(105.5, 57.5){$\big(0,\lambda[3]_1\big)$ }
\put(105.5, -10.5){$\big(0,\lambda[3]_2\big)$}
\put(60.5, 20.5){$\ \ \ + \ \  \ \ \ldots \ldots\ldots\ldots\ldots \ldots\ldots $ }
\end{picture}}
\caption{\label{m0m0m1}Cubic vertex    for  massive
$\Phi^{(3)}_{\mu[s_1], \nu[s_2],\rho[s_3]}$  of spin $s[3]$ and two massless fields $\Phi^{(i)}_{\mu[{\lambda^{(i)}_1}], \nu[{\lambda^{(i)}_2}],\rho[{\lambda^{(i)}_3}]}$  of  helicities $\lambda[3]_i$, $i=1,2$. The terms  in $"\ldots"$ correspond to the auxiliary fields  from $|\chi^{(i)}\rangle_{s[3]_i}$}
\end{figure}
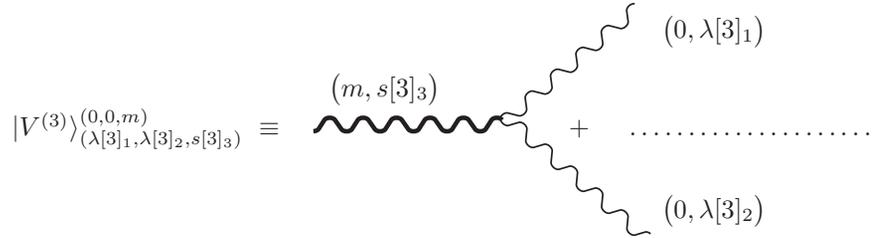

The resolution of    (\ref{g1})  for $x$-local
cubic vertices with representation

\begin{equation}\label{xdep}
  \big|  V^{(3)}_c\rangle^{(m)_{(i)_3}}_{(s[3])_{(i)_3}}  = \prod_{l=1}^3 \delta^{(d)}\big(x -  x_{i_l}\big) V^{(3)(m)_{(i)_3}}_{c|(s[3])_{(i)_3}}(x)
  \prod_{l=1}^3 \eta^{(i_l)}_0 |0\rangle , \ \  |0\rangle\equiv \otimes_{t=1}^p |0\rangle^{t},
\end{equation}
which
(e.g. given by Figure~\ref{m0m0m1} for 2 massless and one massive HS fields) leads to the system
\begin{eqnarray}
\hspace{-0.7em}&\hspace{-0.7em}&\hspace{-0.7em} \mathcal{Q}(V^{(3)l-1}_{c|{(i)_3}},{V}^{(3)l}_{c|{(i)_3}}) = \sum_{n=1}^3
Q^{(i_n)} \big|  {V}^{(3)l-1}_c\rangle
   +  Q^{(i_t)}\Big( \big|  V^{(3)l-1}_c\rangle - \big|  {V}^{(3)l}_c\rangle\Big)=0
\label{g1operV3}   
\end{eqnarray}
 (for $t=1,2,3$  and  $l=0,...,L-1$). \\
 \vspace{10ex}

\noindent
 Its particular solution for coinciding $V^{(3)}_{c|{(i)_3}} = {V}^{(3)0}_{c|{(i)_3}}=...={V}^{(3)k-1}_{c|{(i)_3}}$ has the usual form
but augmented  by the spin,  traceless and Young-symmetry conditions with nilpotent total BRST operator $ Q_c^{tot}$ in space $\otimes_{i=1}^p \mathcal{H}^{(i)}_{tot}$:
\begin{eqnarray}\label{gencubBRSTc}
 &&  Q_c^{tot}
\big|  V^{(3)}_c\rangle^{(m)_{(i)_3}}_{(s[3])_{(i)_3}} =0, \  Q_c^{tot}\equiv \sum_{i=1}^p  Q_c^{(i)},  \qquad    \big( L^{(t)}_{ij},\, T^{(t)}_{rs}\big) \big|  V^{(3)}_c\rangle^{(m)_{(i)_3}}_{({s}[3])_{(i)_3}} \ =\ 0, \\
 && \sigma^{(t)i_t}_c\big|  V^{(3)}_c\rangle^{(m)_{(i)_3}}_{({s}[3])_{(i)_3}}\ =\  \Big(s_{i_t}+\frac{d-2+\theta_{m_i,0}}{2}\Big)\big|  V^{(3)}_c\rangle^{(m)_{(i)_3}}_{({s}[3])_{(i)_3}} .
\label{gencubBRSTc1}
\end{eqnarray}
Without imposing algebraic constraints the deformed LF describes a dynamics for many particles with less spins then for original free HS  fields $\Phi^{(p)}_{\mu[{\hat{s}{}^{(p)}_1}],\nu[{\hat{s}{}^{(p)}_2}], \rho[{\hat{s}{}^{(p)}_3}]}$. Thus, we developed the deformation procedure to get consistent interacting vertices with mixed-antisymmetric HS fields, starting from cubic ones to be explicitly considered  in the subsequent research.

\section{Conclusion}

We developed   BRST approach with complete and incomplete BRST operators to solve new problem of constructing LFs for
mixed-antisymmetric HS field $\Phi_{\mu^1[{\hat{s}_1}],\mu^2[{\hat{s}_2}], \mu^3[{\hat{s}_3}]}$ of generalized integer spin $\mathbf{s} = (3,3,...,3,2,2,...,2,1,...,1)$
in a  Minkowski space $\mathbb{R}^{1,d-1}$ whose symmetry of indices is subject to 3-column Young tableaux $Y[\hat{s}_1, \hat{s}_2, \hat{s}_3]$. The respective LFs for free HS field have equivalent dynamics and are   qualified as Abelian gauge theories of  stages  reducibility depending on sum of spin components  equal respectively to $(\sum_i\hat{s}_i+2)$ and  $(\sum_i\tilde{s}_i-1)$ with fewer auxiliary fields for the model with algebraic constraints.
The results  are applicable both for massless  and massive particles. The Noether deformation procedure is proposed within approach  with incomplete BRST operator  for constructing interacting LF with local cubic vertices for $p$-copies of mixed-antisymmetric   HS fields with preservation the  Poincare group irreps for  deformed (non-Abelian) gauge theory.  The procedure permits to  study interactions with totally-symmetric HS fields. The problem of constructing the BRST-BV (Batalin-Vilkovisky) minimal,  quantum  and effective actions for the interacting LF  may be  considered  following to
extension of BRST-BV approach \cite{2303.02870} with incomplete, $Q_c$ BRST operator  in~\cite{2010.15741}
in order to perturbatively evaluate  average expectation values of the quantities composed from mixed-antisymmetric   HS fields.

\begin{acknowledgments}
We wish to acknowledge the  Organizing Committee of the XXV International Workshop-School QFTHEP'270, E.~Boos  for stimulating discussions and wonderful scientific  spirit. A.R. is thankful to I.~Buchbinder and to M.~Vasiliev for useful comments.
\end{acknowledgments}

\section*{CONFLICT OF INTEREST}

The authors of this work declare that they have no conflicts of interest.


\end{document}